\documentclass[11pt,twoside]{article}
\usepackage{cozumel2005}
\usepackage{epsf}
\usepackage{psfig}
\usepackage{lscape}
\begin{document}
\title{The Vertical Structure of Stars in Edge-on Galaxies}    
\author{Anil C. Seth, Julianne J. Dalcanton and Roelof S. de Jong}   
\affil{U. of Washington, STSci}    

\begin{abstract} 

We present a summary of our recent work on the vertical distribution
of the resolved stellar  
populations in six low mass, edge-on, spiral galaxies observed with
the Hubble Space Telescope Advanced Camera for Surveys (HST/ACS). In
each galaxy we find evidence for an extraplanar stellar component
extending up to 15 scale heights ($\sim$3.5 kpc) above the plane. We
analyze the vertical distribution as a function of stellar age by
tracking changes in the color-magnitude diagram. The young stellar
component ($< 10^8$ yrs) is found to have a scale height larger than
the young component in the Milky Way, suggesting that stars in these
low mass galaxies form in a thicker disk. We also find that the scale
height of a stellar population increases with age, with young main
sequence stars, intermediate age asymptotic giant branch stars, and
old red giant branch stars having succesively larger scale heights in
each galaxy. This systematic trend indicates that disk heating must
play some role in producing the extraplanar stars. We constrain the
rate of disk heating using the observed trend between scale height and
stellar age, and find that the observed heating rates are dramatically
smaller than in the Milky Way. The color distributions of the red
giant branch (RGB) stars well above the midplane indicate that the
extended stellar components we see are moderately metal-poor, with
peak metallicities of [Fe/H]$\sim$-1 and with little or no metallicity
gradient with height. The lack of metallicity gradient can be
explained if a majority of extraplanar RGB stars were formed at early
times and are not dominated by a younger heated population. Our
observations suggest that, like the Milky Way, low mass disk galaxies
also have multiple stellar components.  We examine our results in
light of disk heating and merger scenarios and conclude that both
mechanisms likely played a role in forming the disks of our sample
galaxies.  

\end{abstract}


\section{Introduction}              

The age, morphology and composition of the disk, bulge and halo
components found in typical spiral galaxies all provide significant
constraints on the mechanisms of galaxy formation.  In the Milky Way,
the disk of the galaxy has been found to be a complex object, with a thin disk
whose scale height depends on stellar age and a separate thick disk
component first discovered by \citet{gilmore83} using star counts.
The Milky Way thick disk is dominated by old, moderately metal-poor stars
\citep{wyse95} and has recently been shown to be chemically distinct
from the thin disk \citep[e.g.][]{bensby05}.

Outside the Milky Way, very little detailed information exists about
disk morphologies in other galaxies.  Integrated surface brightness
profiles of edge-on spirals have shown that many galaxies appear to have
thick disks \citep[e.g.][]{burstein79,dalcanton02,pohlen04}, while
recent studies in a handful of nearby edge-on galaxies using resolved
stars have confirmed that extraplanar stars are old and moderately
metal-poor \citep{mould05,tikhonov05a}.  

Many different mechanisms have been used to explain the creation of
thin and thick disks.  These mechanisms fall into three categories:
(1) Creation of thicker components by dynamical heating from a thin
disk by molecular clouds, spiral arms, galaxy interaction etc.
\citep[e.g.][]{spitzer51,lacey91,gnedin03}.  (2) the slow collapse of
the proto-Galaxy forming the thick and thin disk
\citep[e.g.][]{eggen62,gilmore84}, and (3) the
formation of a thick disk from mergers by direct accretion of
stars or by {\it in situ} formation from accreted gas
\citep[e.g.][]{bekki01,gilmore02,abadi03,brook04}.  Testing the
applicability of these theories beyond the Milky Way requires detailed
observations of galaxies with a range of masses and types.  In this
paper we focus on the disks of edge-on, low mass, late type spiral
galaxies. 

The work presented here is a summary of our recently published
results, \citet{seth05a} (Paper~I) and \citet{seth05b} (Paper~II).
A more in depth treatment of this work is presented there.  In \S2 we
present an overview of our galaxy sample and describe out HST/ACS
observations.  We present our primary results on the vertical
distribution of stars and the metallicity of extraplanar stars in our
sample galaxies in \S3.  This is followed by a discussion in \S4.

\section{Observations and Sample}   

We obtained HST/ACS images of 16 edge-on ($i>80^\circ$), late type
galaxies as part of a Cycle~12 snapshot proposal.  A subsample of
eight fields in six galaxies were near enough to enable distance
measurements using the tip of the red giant branch (RGB).  These six
galaxies and their distances, morphological types, circular velocities, $K_s$-band scale
lengths and scale heights\footnote{We use a functional form $\Sigma(z)
  \propto sech^{2}(z/z_{0})$ to fit our surface density profiles.
  Scale heights refer to the $z_0$ in this equation.}
as determined from fits to 2MASS data (Paper~I), and the number of ACS
fields/pointings are listed in Table~1.  Because these galaxies are
nearby, the ACS field-of-view does not cover the entire galaxy.  For
most galaxies the fields are roughly centered on the galaxy,
however, for those galaxies with two fields (NGC~55 and NGC~4631), the
second field (denoted 
with a '-DISK' suffix) is located further out in the disk.  The six
galaxies are all late type (Sc+), none have a significant bulge, and
all have luminosities and inferred masses significantly lower than the
Milky Way.

\begin{table}
\caption{Galaxy Sample Properties}
\smallskip
\begin{center}
{\small
\begin{tabular}{lcccccc}
\tableline \noalign{\smallskip}
Galaxy & Dist. & Type & $V_{max}$ & $K_s$-band $h_r$ & $K_s$-band $z_0$  & \# of ACS \\
  & [Mpc] &  & km/sec & [kpc] & [pc] & Fields \\
\noalign{\smallskip} \tableline \noalign{\smallskip}
\tableline \noalign{\smallskip}
NGC 55      & 2.1 & SBm  & 67  & 0.96 & 437 & 2 \\
IC 5052     & 6.0 & SBcd & 79  & 1.57 & 390 & 1 \\ 
NGC 4144    & 7.5 & SBc  & 67  & 1.10 & 458 & 1 \\ 
NGC 4244    & 4.4 & Sc   & 93  & 1.78 & 257 & 1 \\ 
NGC 4631    & 7.7 & SBcd & 131 & 1.32 & 280 & 2 \\ 
NGC 5023    & 6.6 & Sc   & 77  & 1.28 & 160 & 1 \\ 
\noalign{\smallskip} \tableline \end{tabular} } \end{center}
\end{table}

The ACS observations consisted of 676 seconds of exposure time in the
F606W filter and 700 seconds in the F814W filter.  Our photometric
pipeline is described in detail in Paper~I and yielded 40,000 to
280,000 stars per field.  Extensive artificial star tests ($\sim$5
million stars per field) enabled determination of the completeness as
a function of magnitude, color and local surface brightness allowing
us to study the number of stars as a function of distance from the
disk midplane ('disk height').  To enable adequate correction of the
number counts of stars, we restrict all analysis presented here to
stars brighter than the 20\% completeness level in the highest surface
brightness areas (i.e. the midplane).  

\begin{figure}[!ht]
\plotone{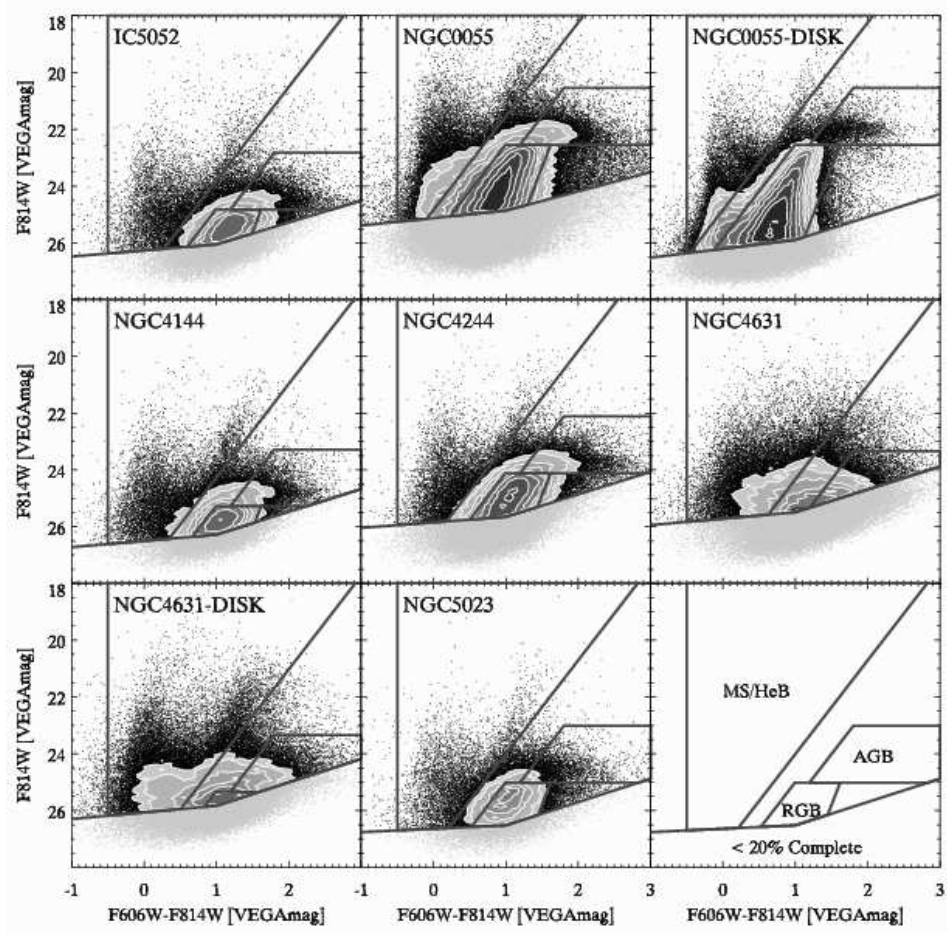}
\caption{Color-magnitude diagrams of each field. Contours are used
  where the density of points becomes high.  The contours are drawn at  
  densities of 75, 100, 150, 200, 250, 350, 500, 750, 1000, and 1500
  stars per 0.1 magnitude and color bin.  The bottom-left panel
  contains a key to the gray lines which demarcate sections of the
  CMD.  The line across the bottom shows the 20\% completeness limit
  in bright regions of the galaxy and the other lines show the regions
  used in each galaxy for the MS/HeB, AGB and RGB stars.}
\end{figure}

Figure~1 shows the color-magnitude diagrams (CMDs) for each of the
fields analyzed here.  In the nearest galaxies, the red giant branch
(RGB), Asymptotic Giant Branch (AGB), upper Main Sequence (MS) and red
and blue Helium burning star sequences are clearly visible.  We use
these different stellar populations to separate the stars we see into
three broad age categories.  Three boxes are shown on each CMD in
Fig.~1 for the RGB, AGB, and MS (including helium burning) stars.  To
determine how well we could 
isolate different stellar ages we generated synthetic CMDs using the
MATCH code \citep{dolphin02} and assuming a constant star formation
rate starting 13 Gyr ago.  The synthetic CMD was then convolved with
the errors obtained from the artificial star tests for each galaxy
(See Paper~II for more details).  The resulting age distribution for
stars falling in the MS, AGB, and RGB boxes in NGC~4144 is shown in Figure~2.
From this figure it is clear that the MS, AGB, and RGB boxes roughly
correspond to young ($\sim100$ Myr), intermediage age ($\sim$1 Gyr),
and old ($\sim10$ Gyr) stellar populations.  However there is some
overlap between the different boxes, particularly between the AGB and
RGB (see Paper~II for more details).

\begin{figure}[!ht]
\plotone{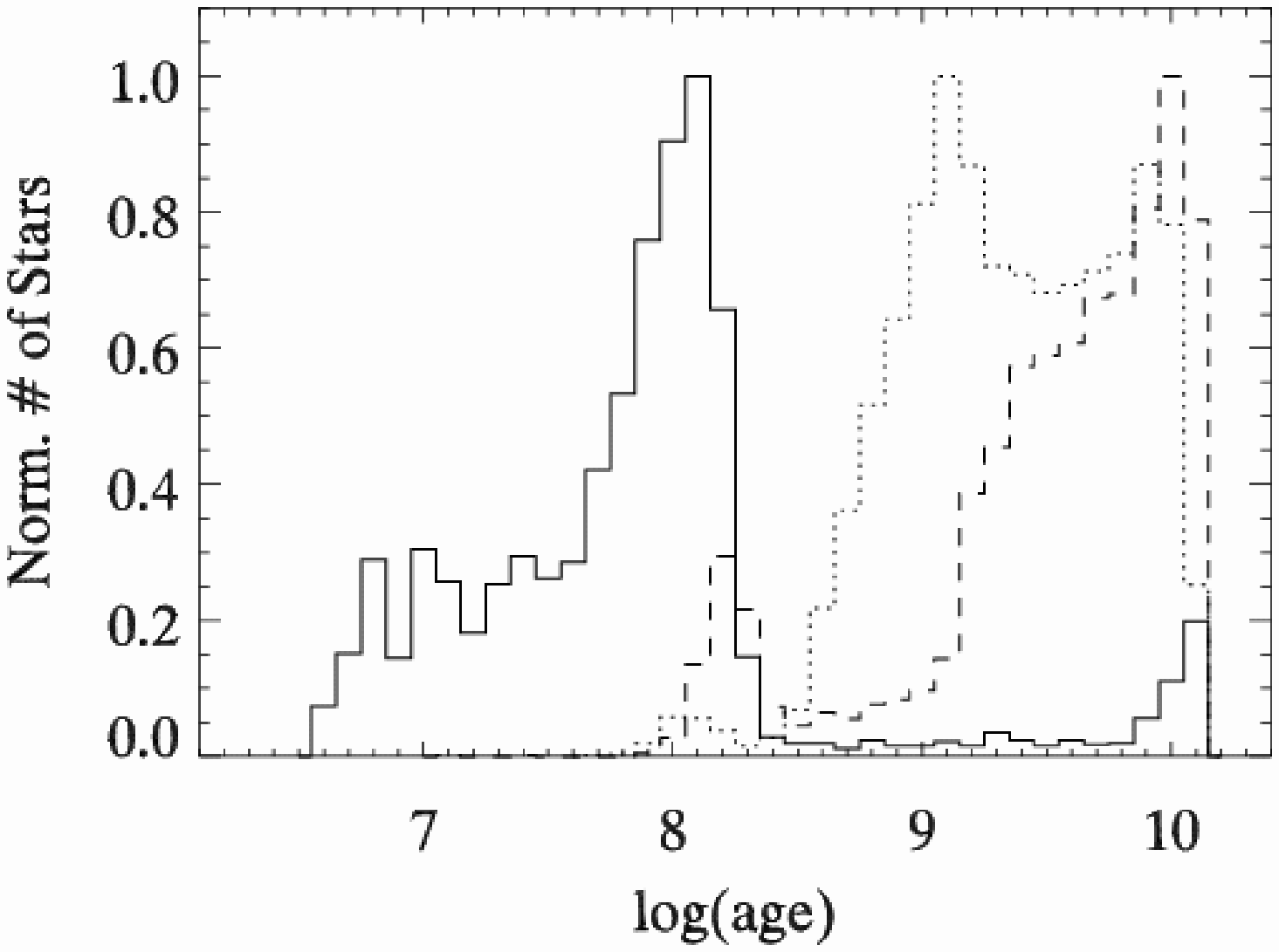}
\caption{Histogram of ages detected in defined CMD boxes (see Fig.~1)
  for NGC~4144 assuming a constant star formation rate from 13 Gyr to
  the present.  Histograms are based on synthesized CMDs created with
  the MATCH program.  This plot shows that stars in the MS box (blue) are
  dominated by stars $\sim$100 Myr in age, while the AGB (red) and RGB
  (orange/yellow) boxes have typical ages of $\sim$1 Gyr and $\sim$10 Gyr
  respectively.  Similar plots for other galaxies are qualitatively
  similar.}
\end{figure}

\section{Results}   

In this section we present the main results of our work.  This
includes an analysis of the scale heights of the resolved stellar
populations and a determination of the metallicity of the old
extraplanar stellar population.

\subsection{Vertical Distribution of Stars}           

Before dividing up the stellar population in different age bins, we
first determine the vertical distribution of all the observed stars
above the 20\% completeness limit.  Figure~3 shows the vertical
stellar density profile for each field in our sample.  This figure was
created by determining the number of stars in
each galaxy in bins at different disk heights.  This number of stars
was then corrected for incompleteness and was divided by the image
area in each bin to obtain a surface density.  The two dashed
lines denote the profiles from the two NGC~4631 fields on the side
where they are contaminated by stars from nearby companion NGC~4627.
The disk heights on the x-axis are scaled by $z_{1/2}$, the disk height
containing 50\% of the $K_s$ band light ($z_{1/2} = 0.5493 z_0$) in
each galaxy.  The two dashed lines show the profile expected for a
stellar population with a scale height 1 and 2 times the $K_s$ band
light.

\begin{figure}[!ht]
\plotfiddle{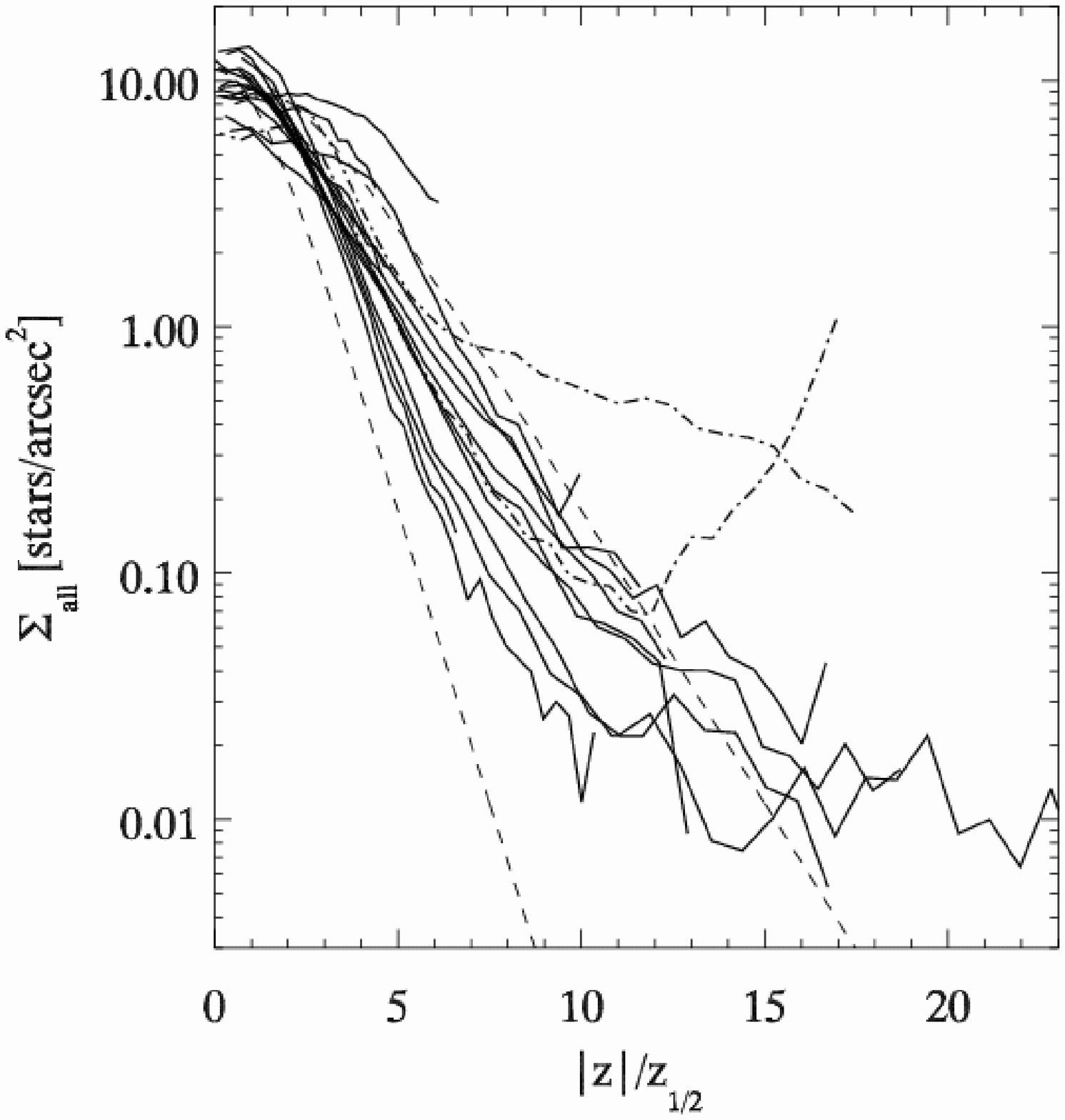}{3.3in}{0}{45}{45}{-144pt}{-45pt}
\caption{The surface density of stars as a function of height from the
  midplane.  Curves have been completeness corrected.  Each galaxy has
  two lines, one for above and below the plane.  The dashed lines
  indicate stellar distributions with scale heights 1 \& 2 $\times$
  that of the $K_s$ band fits to the disks.  The dot-dashed lines
  indicate the side of the NGC~4631 fields where contamination of
  stars from companion NGC~4627 is significant.}
\end{figure}

The most notable result from Figure~3 is that we are able to trace the
resolved stellar population far above the midplane, out to 15-20
$z_{1/2}$ ($\sim$3-4 kpc).  The fact that the star counts continue to
decrease all the way to the edge of the images suggests there are few
background and foreground contaminants.  Assuming a stellar population
at large scale heights similar to Galactic globular clusters
\citep[using data taken from][]{buonanno94,kravtsov97}, we find that we
reach a limiting surface brightness of $\sim$28 mag arcsec$^{-2}$.  We
determine the scale height of the full stellar population at large
disk heights by fitting the profiles between 5 and 10 $z_{1/2}$ above
and below the disk to a
sech$^2$ function.  The derived scale heights listed in Table~2
(in columns 3 \& 4) are typically $\sim$2 times the 2MASS $K_s$ band
scale heights.  We note that the 2MASS images are not very deep and
only trace the highest surface brightness population near the
midplane.  This therefore suggests that the population at large disk
heights has a larger scale height than the population near the midplane.

\begin{table}
\caption{Resolved Stellar Component Scale Heights}
\smallskip
\begin{center}
{\small
\begin{tabular}{lccccccc}
\tableline \noalign{\smallskip}
Field & $K_s$ & All $+$ & All $-$ & MS & AGB & RGB & $h_{r}/z_{\rm
  0,RGB}$ \\
 & $z_0$ [pc] & $z_0$ [pc] & $z_0$ [pc] & $z_0$ [pc] & $z_0$ [pc] &
$z_0$ [pc] & \\
\noalign{\smallskip}
\tableline \noalign{\smallskip}
\tableline \noalign{\smallskip}
\noalign{\smallskip}
IC 5052      & 390 & 767$\pm$30 & 686$\pm$44 &  261$\pm$11 & 467$\pm$6  & 655$\pm$6    & 2.4 \\ 
NGC 55                        & 437 &            &            &  327$\pm$24 & 644$\pm$10 & 701$\pm$3    & 1.4 \\ 
NGC 55-DISK                   & 437 &            &            &  526$\pm$1  & 741$\pm$15 & 999$\pm$6    & 1.0 \\ 
NGC 4144                      & 457 & 940$\pm$41 & 965$\pm$15 &  374$\pm$27 & 699$\pm$16 & 934$\pm$18   & 1.2 \\ 
NGC 4244                      & 468 & 740$\pm$40 &            &  325$\pm$20 & 443$\pm$24 & 551$\pm$9    & 3.2 \\ 
NGC 4631     & 510 & 927$\pm$46 &            &  510$\pm$26 & 895$\pm$51 & 1154$\pm$194 & 1.1 \\ 
NGC 4631-DISK & 510 &1131$\pm$50 &            &  505$\pm$22 & 1200$\pm$1 & 1387$\pm$73  & 1.0 \\ 
NGC 5023     & 291 & 505$\pm$32 & 534$\pm$26 &  204$\pm$6  & 289$\pm$6  & 391$\pm$4    & 3.3 \\ 
\noalign{\smallskip}
\tableline \noalign{\smallskip}
\end{tabular} } \end{center}
\end{table}

We now analyze the scale height of our sample of galaxies as a
function of stellar age.  Using the MS, AGB, and RGB boxes shown in
Figure~1, we are able to isolate young, intermediate, and old stellar
populations.  Figure~4 shows the completeness corrected surface
density profiles for the three different stellar population boxes in
each field.  The youngest MS stars
are shown with a solid line, while the intermediate-age AGB and old RGB
stars are shown with dotted and dashed lines, respectively. The
stellar density profiles have all been normalized to integrate to one.
Each field clearly shows that the MS component is the narrowest while
the RGB is in all cases the thickest component.  We quantified this
trend by fitting a sech$^2$ function to each component excluding the
central portions of the profile from the fit.  The central portions
of the profiles show dips in the star counts which likely result from
dust absorption (see Paper~II).  The scale heights found from the
sech$^2$ fits are shown in columns 5-7 of Table~2.  A number of
interesting trends are visible.  First, in each galaxy/field, the MS
component has the smallest scale height with the AGB and RGB having
succesively larger scale heights.  In other words, the scale height
increases with increasing stellar age in all six galaxies.  Second,
the scale heights of the MS component are larger than (200-500 pc) than the
scale height of young stars in the Milky Way \citep[$\sim$200
  pc;][]{schmidt63} despite being less luminous and having shorter
scale lengths.  This thicker MS component may be related to the thicker
dust distribution seen in low mass galaxies by \citet{dalcanton04}.
Finally, the axial ratio of the RGB component ($h_{r}/z_{\rm 0,RGB}$)
shown in the last column of Table~2 ranges from 1.0 to 3.3.  This
axial ratio is similar to the Milky Way thick disk
\citep[using data from e.g][]{chen01}. 

\begin{figure}[!ht]
\plotone{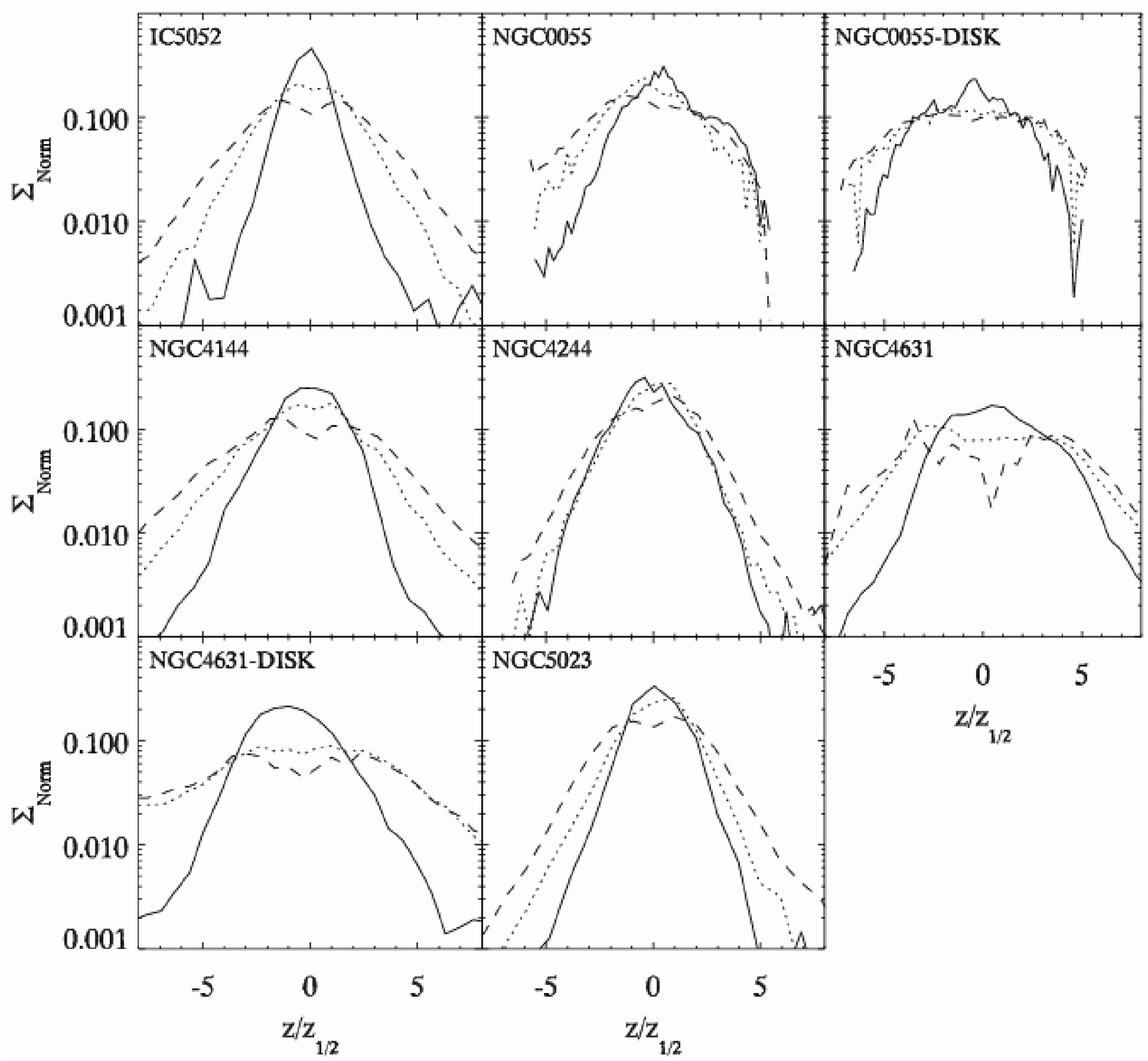}
\caption{The normalized surface density as a function of scaleheight
  for young MS (solid), intermediate-age AGB (dotted) and old RGB
  (dashed) stars.  Each surface density distribution was normalized to
  integrate to one.  Note that in all cases the MS distribution is the
  most peaked while the RGB distribution is the widest.}
\end{figure}

\subsection{Metallicity}           

Based on the scale height results, we expect most of the stars at
large disk heights to be old.  The morphology of the RGB population
can provide a metallicity diagnostic for old stellar populations
\citep[e.g.][]{dacosta90,armandroff93,frayn02}.  In this section we
use the color of the RGB stars to determine the metallicity of the
extraplanar stellar population.

The left panel of Figure~5 shows the composite CMD of stars above
4$z_{1/2}$ in all the galaxies excepting the NGC~4631 fields (which
have completeness limits that are too high for RGB metallicity
analysis).  Overplotted are 10 Gyr Padova isochrones at [Fe/H] values 
ranging between -2.3 and 0.0.  The peak of the distribution
lies between the isochrones with [Fe/H] of -1.3 and -0.7 indicating
that the dominant population in these galaxies at high latitudes is
moderately metal-poor.  We determined a metallicity distribution
function for each galaxy using three bins in absolute
magnitude shown with dotted lines in the left panel of Figure~5.  In
each bin we determined the metallicity of individual stars by
linearly interpolating between the 10 Gyr Padova isochrones.  The
right panel of Figure~5 shows the metallicity distribution 
functions for NGC~4244, with the shaded grey region indicating the
range of results for the three different bins.  The peak metallicity
is at [Fe/H] of -0.9 with a long tail towards lower metallicities and
a sharper cutoff at higher metallicities.  The other four galaxies
have similar MDFs, with peaks ranging between [Fe/H] of -0.7 and -1.1.
This peak metallicity also agrees with metallicities found for
extraplanar stars in other studies of galaxies in the same mass range
\citep{cole00,davidge03,tiede04,davidge05}.  
This metallicity is slightly more metal-poor than the Milky Way thick disk
metallicity ([Fe/H]$\sim$-0.6) but significantly more metal-rich than
the Milky Way halo \citep[${\rm [Fe/H] \sim}$-1.7][]{wyse95}.  We note that
the current metallicity of these galaxies is somewhat subsolar
\citep{garnett02}, furthering the idea that the peak metallicites of
[Fe/H]$\sim$-1 are consistent with the presence of a thick disk
similar to the Milky Way's.

\begin{figure}
\plottwo{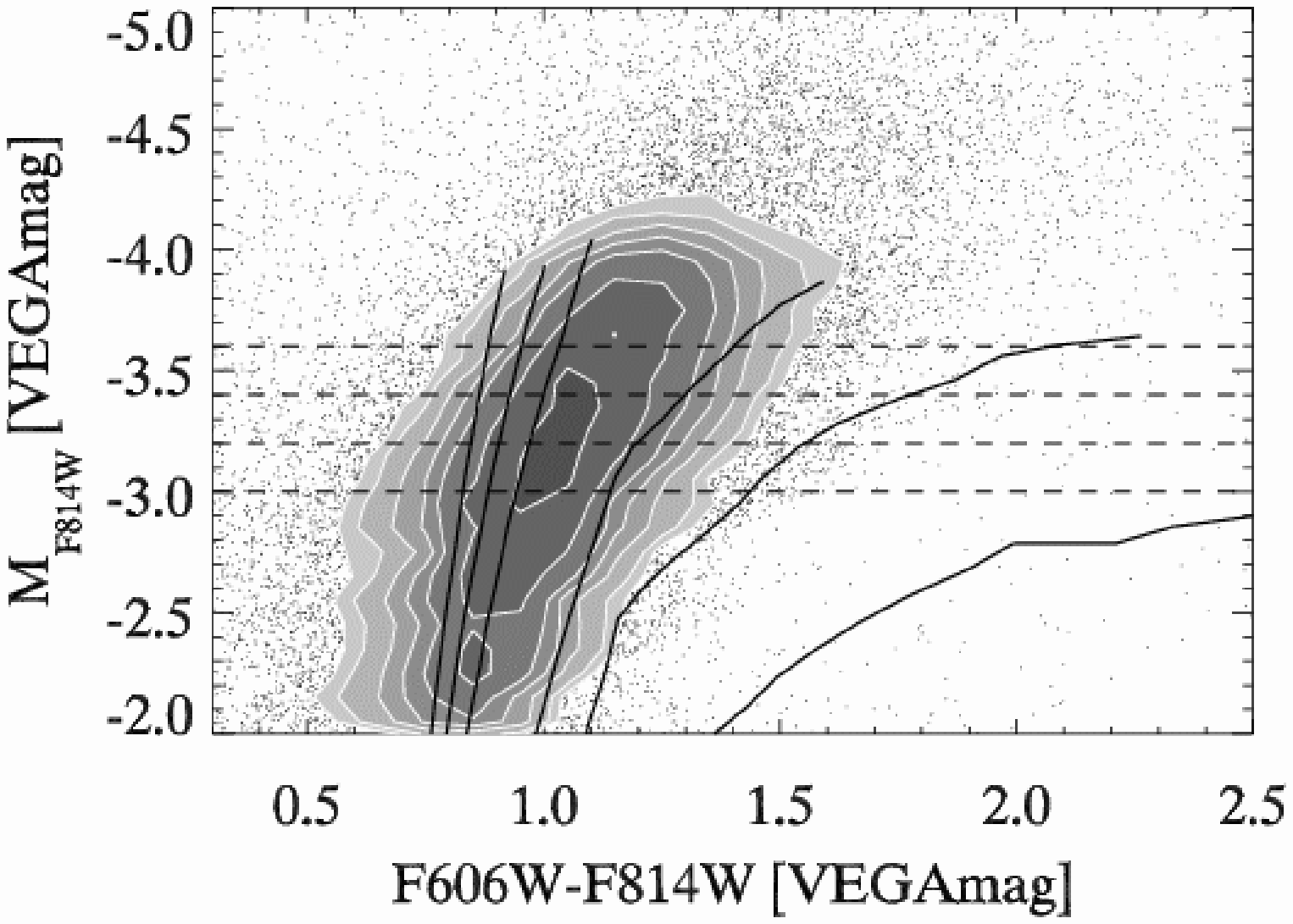}{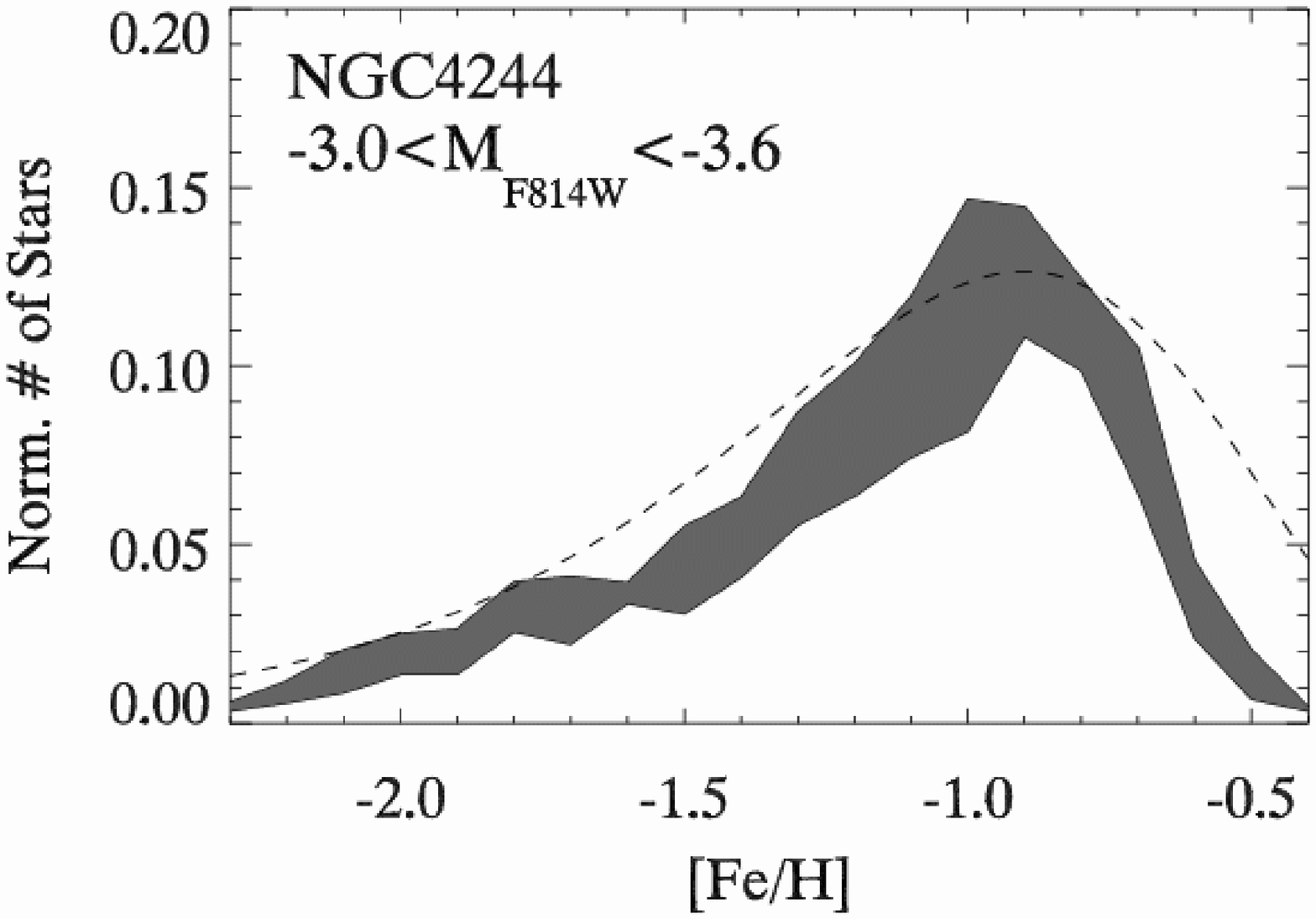}

\caption{{\it (left) -- } Composite CMD of extraplanar stars in
  IC~5052, NGC~55, NGC~55-DISK, NGC~4144, NGC~4244, and NGC~5023 with
  disk heights $>$4$z_{1/2}$.  Solid lines show 10 Gyr old Padova
  models for the RGB with [Fe/H] (from left to right) of -2.3, -1.7,
  -1.3, -0.7, -0.4, and 0.0.  The peak of the stellar distributions
  fall between [Fe/H] of -1.3 and -0.7.  {\it (right) -- } The
  metallicity distribution function for stars at disk heights
  $>$4$z_{1/2}$ in NGC~4244.  Gray regions indicate the range of
  values obtained in the different absolute magnitude bins shown with
  dashed lines in the left panel.  The dashed line in this panel indicates a
  closed-box chemical evolution models scaled to the peak of the MDF.}

\end{figure}

Overplotted in the right panel of Figure~5 is a closed-box ``simple''
chemical evolution model \citep[Eq. 20 of][]{pagel97} scaled to the
peak of the MDF.  The overall shape is similar to observed MDF,
however our data appears to have a sharper cutoff at the high
metallicity end, and a 
deficit of stars at the low metallicity end.  The high metallicity
cutoff suggests that the star formation trucated before all the gas
was consumed, as would be expected for any of the thick disk scenarios
discussed in the introduction.  At low metallicities, the deficit of
stars may be a manifestation of the G-dwarf problem.  However, our results at
the low metallicity end are quite uncertain since the close spacing of
the isochrones means photometric errors introduce large uncertainties
in the metallicity.  

Models for the origin of thin and thick disks predict different degrees of
variation in the stellar metallicity with height above the plane.  We
can observe any such trend in our galaxies by examining the median
color of the RGB stars as a function of height above the midplane.
Figure~6 shows the median RGB color vs. disk height for each of the
galaxies in our sample (except NGC~4631).  At low disk heights,
internal reddening may impact the color of stars - this is shown as
the hatched region.  Above this height, from $\sim$3-10$z_{1/2}$
($\sim$1-2 kpc), the galaxies show very little variation in RGB color
suggesting nearly uniform metallicity with increasing disk height.
Recent observations by \citep{davidge05,tikhonov05a,mould05} of
extraplanar stars in similar galaxies also find a lack of strong
metallicity gradients.  We discuss the implications of this further in
the next section.

\begin{figure}[!ht]
\plotone{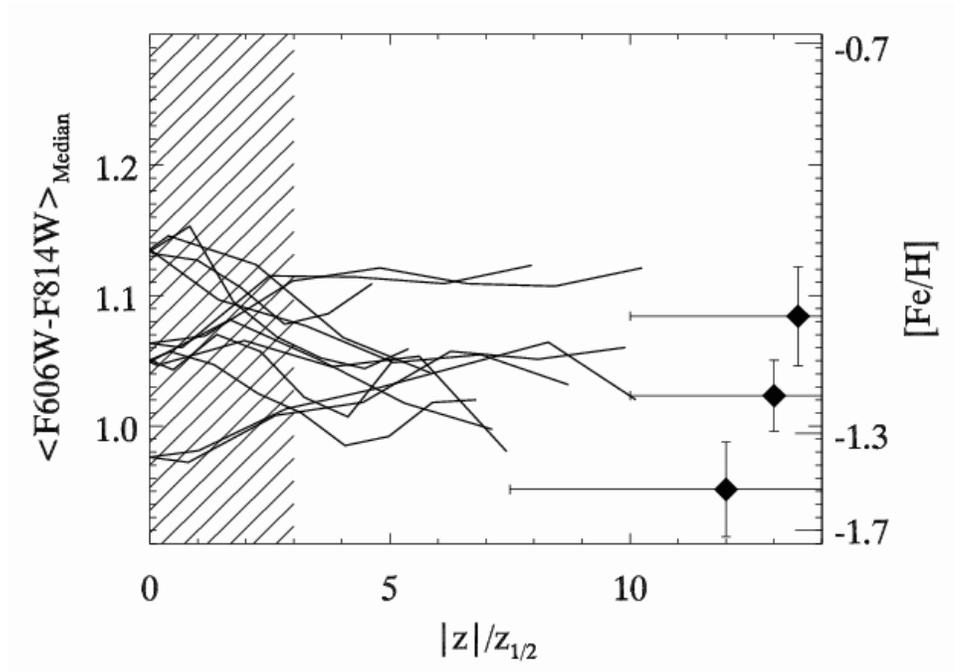}
\caption{The median RGB color as a function of disk height in each
  galaxy.  This includes stars with F606W-F814W redder than 0.7 and
  M$_{\rm F814W}$ between -3.2 and -3.6. The error on all plotted data
  is $<$0.05 magnitudes.  The diamonds indicate the median color of
  all stars at large disk heights and are plotted only for the three
  galaxies with significant numbers of stars at large scale heights.
  The hatched area indicates the heights at which internal reddening
  may affect the stellar colors.  The right axis shows the mean RGB
  color of Padua isochrones \citep{girardi04} at three different
  metallicities.}
\end{figure}

\section{Discussion and Conclusions}

We have shown above that in our sample of six edge-on, low mass, late
type galaxies:
\begin{itemize}
\item Stars exist at large heights above the disk plane.
\item The scale height of a stellar population increases with age in
 each galaxy.  The young (MS) stellar population has a larger scale height
 than young stars in the Milky Way, while the old RGB population has
 an axial ratio similar to the Milky Way thick disk.
\item Extraplanar RGB stars have a peak metallicity of
  [Fe/H]$\sim$-1.  In addition there appears to be little gradient in
  the metallicity at moderate (1-2 kpc) disk heights.
\end{itemize}

In the Milky Way, it is widely accepted that the age-dependent scale
height in the thin disk is a result of slow disk heating, while the
thick disk is thought to be formed by some type of merger event.  We
now consider our results in light of this scenario.

\vspace{0.1in}

\noindent {\it Disk Heating: }  The steady increase in scale height
with stellar population age suggests that some kind of disk heating is
occurring in our sample galaxies.  We can use our measured scale heights 
to constrain the rate of disk heating in these galaxies.  For an
isothermal sech$^2$ profile, the scale height of a stellar population
is proportional to the square of the vertical velocity dispersion
($z_0 \propto \sigma_{z}^{2}$).  In the Milky Way, the rate of disk
heating is usually expressed as a power law, $\sigma_z \propto
t^{-\alpha}$, with $\alpha$ ranging between 0.3 and 0.6
\citep{hanninen02}.  Using the ages for our stellar population boxes
derived from the constant star-formation rate synthetic CMDs, we find
that the rate of disk heating is much lower, with $\alpha$ being no
larger than 0.15 (see Paper~II for more details).  This is not
surprising given our galaxies likely have fewer molecular clouds and
weaker spiral arms.  Based on this slow heating rate, it is plausible
that disk heating can account for the observed variation of scale
height.  However, a model in which steady disk heating accounts for all of the
extraplanar stars would predict the existence of intermediate-age RGB
stars with higher metallicities and lower scale heights leading to a
decreasing metallicity with increasing disk height.  The lack of such
a trend in our galaxies suggests that a majority of RGB stars at all
disk heights formed early and with a well-mixed metallicity
distribution.  This strongly suggests that mergers or interactions
played a role in the formation of these RGB stars.

\vspace{0.1in}

\noindent {\it Merger \& Interactions: } The thick disk in the Milky
Way is thought to have formed in a merger event
\citep[e.g.]{abadi03,brook04}, and recent dynamical observations of a
counter-rotating thick disk in FGC~227 by \citet{yoachim05} also strongly
suggests a merger origin.  In our sample galaxies, the lack of
metallicity 
gradient and the similarity of the axial ratio and metallicity of the
RGB population to the Milky Way thick disk suggests that mergers or
interactions likely have played an important role in the early history
of these galaxies.  Further evidence to support this claim comes from
the heterogeneity of scale heights we observe.  Although all the
galaxies in our sample show a similar trend towards increasing scale
height with age, in some galaxies the RGB component is significantly
thicker than in others despite having similar circular velocity/mass.
This variation would be a natural consequence of the stochastic
merging process.

Overall, our results require that some disk heating occurs at
intermediate ages (to puff up the AGB component), but that events at
earlier times (interactions or mergers) created a majority of the RGB
stars over a short period of time.  In future work, we will focus on the
globular cluster systems of our sample galaxies which will provide us with
additional evidence on their history and assembly.

\acknowledgements 
The authors would like to thank Andrew Dolphin and
Antonio Aparicio for their help in generating synthetic CMDs and Leo
Girardi for supplying us with isochrones.  This work was supported by
HST-GO-09765, the Sloan foundation, and NSF Grant CAREER AST-0238683.



\end{document}